\documentclass[fleqn,twoside]{article}
\usepackage{espcrc2}

\usepackage{graphicx}

\newcommand{\bdu}{\mbox{\boldmath $u$}}

\title{Basis Optimization Renormalization Group for Quantum Hamiltonian}

\author{Takanori Sugihara
\thanks{Email: sugihara@eken.phys.nagoya-u.ac.jp}\\
\vspace{0.2cm}
{Department of Physics, Nagoya University, \\ 
Chikusa, Nagoya 464-8602, Japan}}
\begin{document}

\begin{abstract}
We find an algorithm of numerical renormalization group for 
spin chain models. 
The essence of this algorithm is orthogonal transformation 
of basis states, which is useful for reducing the number of 
relevant basis states to create effective Hamiltonian. 
We define two types of rotations and combine them to create 
appropriate orthogonal transformation. 
\vspace{1pc}
\end{abstract}

\maketitle

\section{INTRODUCTION}
A method of Hamiltonian diagonalization is suitable for 
calculating physical quantities associated with wavefunctions, 
such as structure functions and form factors. 
However, we cannot diagonalize Hamiltonian directly 
in quantum field theories because dimension of Hilbert space 
is generally infinite. 
To create effective Hamiltonian, we extend a technique 
of NRG (numerical renormalization group) 
proposed by K. Wilson \cite{wilson}. 

In spin chain models, 
any state can be expanded with a direct product 
of two sets of basis states, each of which 
is for one of two spin blocks. 
For a finite lattice, we can calculate wavefunction $\Psi_{ij}$ 
of the ground state by diagonalizing Hamiltonian. 
In Ref. \cite{white}, S.White has found that 
singular values $D_k$ of target-state wavefunction 
$\Psi_{ij}=\sum_k U_{ik}D_k V_{kj}$ are useful for 
the purpose of creating effective Hamiltonian. 
\begin{equation}
  |\Psi\rangle = \sum_i \sum_j \Psi_{ij}|i\rangle |j\rangle
  = \sum_k D_k |u_k\rangle |v_k\rangle, 
\end{equation}
where 
$|u_k\rangle\equiv\sum_i U_{ik}|i\rangle$ and 
$|v_k\rangle\equiv\sum_j V_{kj}|j\rangle$. 
If $|D_k|$ is large, the corresponding basis state is relevant. 
However, if $|D_k|$ is small, the corresponding 
basis state is not relevant and can be thrown away from calculation. 
We can control calculation accuracy using singular values $D_k$ 
when we truncate Hilbert space. 
Magnitude of $|D_k|$ determines the relevance of basis states 
in the target state. Basis truncation based on singular values 
works well in one-dimensional spin models such as Heisenberg 
and Hubbard models \cite{white,nw}. This is an extension of 
Wilson's NRG and called DMRG (density matrix renormalization group). 
White has shown that there exists orthogonal transformation 
that largely reduces the number of basis states 
with high calculation accuracy. 
However, we cannot assure that a number of singular values take 
values of nearly zero in general cases. 
In order to create optimized basis states without relying on 
density matrix, we will find another useful basis transformation. 
Our aim is to find basis transformation that make coefficients 
(i.e. transformed wavefunctions) exactly zero 
when target states are expanded with new basis states. 

\section{FORMULATION}
Let us consider $S=1/2$ Heisenberg chain. 
Since we cannot diagonalize infinite dimensional Hamiltonian directly, 
we start from a finite lattice. We extend a lattice step by step 
by adding one site to it in each RG transformation. 
$|u_i\rangle$ is a complete set of basis states for a finite lattice $L$. 
Hilbert space of an extended lattice with size ($L+1$) is 
spanned with a set of basis states $|u_i\rangle |j\rangle$, 
where $|j\rangle$ is an eigenstate of $z$-component of 
a spin operator for the ($L+1$)-th site 
\[
\hat{S}^z|j\rangle = \pm \frac{1}{2}|j\rangle
\quad \mbox{for}\quad j=\pm. 
\]
We want to find a rotation matrix $U$ 
that decrease the number of relevant basis states. 
\begin{equation}
  |u'_j\rangle = \sum_i U_{ij}|u_i\rangle. 
\end{equation}
In general, any orthogonal matrix can be expressed 
as a product of two-dimensional rotation matrices $R$. 
\[
  R_{(k,l)}(\theta)=
  \bordermatrix{
    &        & k          &        & l           &        \cr
    &        & \vdots     &        & \vdots      &        \cr
  k & \ldots & \cos\theta & \ldots & -\sin\theta & \ldots \cr
    &        & \vdots     & \ddots & \vdots      &        \cr
  l & \ldots & \sin\theta & \ldots & \cos\theta  & \ldots \cr
    &        & \vdots     &        & \vdots      &        \cr}
\]
This matrix rotates the $k$-th and $l$-th basis states by 
an angle $\theta$. 
Let us consider a ground state that is an eigenstate of a Hamiltonian. 
\begin{equation}
  |\Psi\rangle = \sum_{ij} \Psi_{ij} |u_i\rangle |j\rangle. 
\end{equation}
We consider a part of this target state. 
\begin{eqnarray*}
  &&|\Psi\rangle_{(1,2)} =
  \Psi_{1-}|u_1\rangle|-\rangle + \Psi_{1+}|u_1\rangle|+\rangle
  \\
  &&\hspace{1.3cm}
  +\Psi_{2-}|u_2\rangle |-\rangle + \Psi_{2+}|u_2\rangle|+\rangle. 
\end{eqnarray*}
If we rotate two basis states $|u_1\rangle$ and $|u_2\rangle$ 
\begin{eqnarray*}
  &&|u_1\rangle =
  \cos\theta|u'_1\rangle-\sin\theta|u'_2\rangle, 
  \\
  &&|u_2\rangle =
  \sin\theta|u'_1\rangle+\cos\theta|u'_2\rangle, 
\end{eqnarray*}
we have 
\begin{eqnarray*}
  &&|\Psi\rangle_{(1,2)} =
  \Psi'_{1-}|u'_1\rangle|-\rangle + \Psi'_{1+}|u'_1\rangle|+\rangle
  \\
  &&\hspace{1.3cm}
  +\Psi'_{2-}|u'_2\rangle |-\rangle + \Psi'_{2+}|u'_2\rangle|+\rangle, 
\end{eqnarray*}
where $\Psi'$ are transformed wavefunctions 
\begin{eqnarray*}
  &&\Psi'_{1-} = \Psi_{1-}\cos\theta + \Psi_{2-}\sin\theta, 
  \\
  &&\Psi'_{1+} = \Psi_{1+}\cos\theta + \Psi_{2+}\sin\theta, 
  \\
  &&\Psi'_{2-} =-\Psi_{1-}\sin\theta + \Psi_{2-}\cos\theta, 
  \\
  &&\Psi'_{2+} =-\Psi_{1+}\sin\theta + \Psi_{2+}\cos\theta. 
\end{eqnarray*}
We define the following two types of rotations. 
Rotation I is defined to make the fourth component zero $\Psi'_{2+}=0$; 
\[
  \pmatrix{
  \Psi_{1-} \cr
  \Psi_{1+} \cr
  \Psi_{2-} \cr
  \Psi_{2+} }
  \stackrel{\rm I}{\to}
  \pmatrix{
  \Psi_{1-}' \cr
  \Psi_{1+}' \cr
  \Psi_{2-}' \cr
  0 \cr}
\]
A condition $\Psi_{2+}'=0$ determines the value of $\theta$. 
Rotation II is defined to make the third component zero $\Psi_{2-}=0$ 
when components $\Psi_{1+}$ and $\Psi_{2+}$ are zero; 
\[
  \pmatrix{
  \Psi_{1-} \cr
  0 \cr
  \Psi_{2-} \cr
  0 }
  \stackrel{\rm II}{\to}
  \pmatrix{
  \Psi_{1-}' \cr
  0 \cr
  0 \cr
  0 \cr}
\]
When $\Psi_{1+}=\Psi_{2+}=0$, the transformed components 
$\Psi'_{1+}$ and $\Psi'_{2+}$ are always zero. 
Therefore, we can make $\Psi'_{2-}$ zero 
by choosing $\theta$ appropriately. 

In order to understand how these two rotations are used to 
find relevant basis states, 
let us consider a simple example where a target state 
is expanded with six basis states. 
\begin{equation}
  |\Psi\rangle =
  \sum_{i=1}^3 \sum_{j=\pm}
  \Psi_{ij} |u_i\rangle |j\rangle.
  \label{sixdim}
\end{equation}
The following is transformation of the six-dimensional vector 
of the target-state wavefunction $\Psi_{ij}$. 
  \[
  \pmatrix{
  \Psi_{1-} \cr
  \Psi_{1+} \cr
  \Psi_{2-} \cr
  \Psi_{2+} \cr
  \Psi_{3-} \cr
  \Psi_{3+} }
  \stackrel{\rm I}{\to}
  \pmatrix{
  \Psi_{1-}' \cr
  \Psi_{1+}' \cr
  \Psi_{2-}' \cr
  0          \cr
  \Psi_{3-} \cr
  \Psi_{3+} }
  \stackrel{\rm I}{\to}
  \pmatrix{
  \Psi_{1-}'' \cr
  \Psi_{1+}'' \cr
  \Psi_{2-}'  \cr
  0           \cr
  \Psi_{3-}'  \cr
  0          }
  \stackrel{\rm II}{\to}
  \pmatrix{
  \Psi_{1-}'' \cr
  \Psi_{1+}'' \cr
  \Psi_{2-}'' \cr
  0           \cr
  0           \cr
  0           }
\]
We obtain the final form by transforming the vector 
with two rotations appropriately. 
If a transformed wavefunction is zero, the corresponding basis state 
is irrelevant and can be neglected. That is, only the first three 
basis states 
($|u_1''\rangle|\pm\rangle$, $|u_2''\rangle|-\rangle$) 
are relevant and others 
($|u_2''\rangle|+\rangle$, $|u_3''\rangle|\pm\rangle$) 
are redundant. 
The former three relevant basis states can reproduce 
the target state exactly without other basis states. 
There exists orthogonal transformation that reduces 
the number of relevant basis states into {\it three} 
without causing calculation error. 
This is also true for arbitrary dimensions 
other than six of this example. 

Let us consider a case where we target ground and first-excited 
states. The followings are wavefunction vectors of the target states. 
\[
  \pmatrix{
  \Psi_{1-}^{(1)} &\!\!\!\!\!\! \Psi_{1-}^{(2)} \cr
  \Psi_{1+}^{(1)} &\!\!\!\!\!\! \Psi_{1+}^{(2)} \cr
  \Psi_{2-}^{(1)} &\!\!\!\!\!\! \Psi_{2-}^{(2)} \cr
  \Psi_{2+}^{(1)} &\!\!\!\!\!\! \Psi_{2+}^{(2)} \cr
  \Psi_{3-}^{(1)} &\!\!\!\!\!\! \Psi_{3-}^{(2)} \cr
  \Psi_{3+}^{(1)} &\!\!\!\!\!\! \Psi_{3+}^{(2)} \cr
  \Psi_{4-}^{(1)} &\!\!\!\!\!\! \Psi_{4-}^{(2)} \cr
  \Psi_{4+}^{(1)} &\!\!\!\!\!\! \Psi_{4+}^{(2)} \cr
  \vdots          &\!\!\!\!\!\! \vdots          \cr
  \Psi_{N-}^{(1)} &\!\!\!\!\!\! \Psi_{N-}^{(2)} \cr
  \Psi_{N+}^{(1)} &\!\!\!\!\!\! \Psi_{N+}^{(2)} }
  \to
  \pmatrix{
  \Psi_{1-}^{(1)'} &\!\!\!\!\!\! \Psi_{1-}^{(2)'} \cr
  \Psi_{1+}^{(1)'} &\!\!\!\!\!\! \Psi_{1+}^{(2)'} \cr
  \Psi_{2-}^{(1)'} &\!\!\!\!\!\! \Psi_{2-}^{(2)'} \cr
  0                &\!\!\!\!\!\! \Psi_{2+}^{(2)'} \cr
  0        &\!\!\!\!\!\! \Psi_{3-}^{(2)'} \cr
  0        &\!\!\!\!\!\! \Psi_{3+}^{(2)'} \cr
  0        &\!\!\!\!\!\! \Psi_{4-}^{(2)'} \cr
  0        &\!\!\!\!\!\! \Psi_{4+}^{(2)'} \cr
  \vdots   &\!\!\!\!\!\! \vdots           \cr
  0        &\!\!\!\!\!\! \Psi_{N-}^{(2)'} \cr
  0        &\!\!\!\!\!\! \Psi_{N+}^{(2)'} }
  \to
  \pmatrix{
  \Psi_{1-}^{(1)'} & \!\!\!\!\!\! \Psi_{1-}^{(2)'} \cr
  \Psi_{1+}^{(1)'} & \!\!\!\!\!\! \Psi_{1+}^{(2)'} \cr
  \Psi_{2-}^{(1)'} & \!\!\!\!\!\! \Psi_{2-}^{(2)'} \cr
  0                & \!\!\!\!\!\! \Psi_{2+}^{(2)'} \cr
  0                & \!\!\!\!\!\! \Psi_{3-}^{(2)''} \cr
  0                & \!\!\!\!\!\! \Psi_{3+}^{(2)''} \cr
  0                & \!\!\!\!\!\! \Psi_{4-}^{(2)''} \cr
  0                & \!\!\!\!\!\! 0                 \cr
  \vdots           & \!\!\!\!\!\! \vdots            \cr
  0                & \!\!\!\!\!\! 0                 \cr
  0                & \!\!\!\!\!\! 0                 }
\]
The indices $(1)$ and $(2)$ indicate the ground and 
first-excited states, respectively. 
We optimize basis states in two steps. 
In the first step, we target the ground state $\Psi^{(1)}$ 
and do not take care of the first excited state $\Psi^{(2)}$. 
In the same way as the previous example (\ref{sixdim}), 
we can find three relevant 
basis states that can reproduce the ground state exactly 
by using a combination of rotations I and II. 
In the second step, we target the first excited state $\Psi^{(2)}$ 
to optimize basis states further. 
When we rotate basis states, 
we don't touch the first four basis states 
($|u_1'\rangle|\pm\rangle$, $|u_2'\rangle|\pm\rangle$). 
We apply a combination of rotations I and II 
to other ($2N-4$) pieces of basis states 
($|u_3'\rangle|\pm\rangle$, $|u_4'\rangle|\pm\rangle$, $\dots$, 
$|u_N'\rangle|\pm\rangle$). 
Finally, we obtain seven basis states 
($|u_1''\rangle|\pm\rangle$, $|u_2''\rangle|\pm\rangle$, 
$|u_3''\rangle|\pm\rangle$, $|u_4''\rangle|-\rangle$) 
that are relevant for reproducing 
the ground and first excited states exactly. 
In the same way, we can also create relevant basis states 
when arbitrary number of eigenstates are targeted. 
The number of target states can be chosen 
according to the desired calculation accuracy. 
We can use this technique of basis transformation 
to truncate Hilbert space and create effective Hamiltonian in NRG. 
Our NRG algorithm for $S=1/2$ Heisenberg chain is as follows: 
\begin{enumerate}
\item Diagonalize $H_L$ ($H_L \bdu_n = E_n \bdu_n$), 
which is Hamiltonian for a finite lattice $L$. 
\item Transform basis states $\bdu_n\to\bdu_n'$ 
using type I and II rotations to make wavefunctions 
of target states zero as many as possible. 
\item Create effective Hamiltonian $\bar{H}_L=O^\dagger_L H_L O_L$, 
where $O_L=(\bdu_1',\bdu_2',\dots,\bdu_M')$. 
Dimension of $\bar{H}_L$ is $M$.
\item Add one site to $\bar{H}_L$ to provide next Hamiltonian $H_{L+1}$.
Dimension of $H_{L+1}$ is $2M$.
\item Regard $H_{L+1}$ as $H_L$, and go to step 1.
\end{enumerate}
These steps are iterated till vacuum energy per site 
converges with the desired accuracy. 
I have applied this algorithm to $S=1/2$ Heisenberg chain 
and calculated vacuum energy per site. 
An exact value of vacuum energy has been obtained using Bethe ansatz. 
The difference between our result and exact value is about 
$3.9\times 10^{-4}$ for parameters $M=256$ and $L=2000$.

\section{CONCLUSION}
We have found basis transformation that largely reduces 
the number of relevant basis states. 
Especially, three states are sufficient for reproducing 
one target state in each RG step. 
Relevant basis states reproduce target states exactly in each RG step. 
Therefore, if dimension of effective Hamiltonian $M$ is 
sufficiently large, calculation errors come from non-targeted states. 
Future problems are improvement of accuracy, 
calculation of correlation functions, calculation of Haldane gap, 
and application to other models.

\end{document}